\documentclass[journal,onecolumn]{IEEEtran}
\usepackage{amsmath,graphicx}
\usepackage{epstopdf}
\usepackage{marginnote}
\usepackage{pseudocode}
\usepackage[font=normal]{subfig}
\usepackage{graphicx}
\usepackage{tabularx}
\usepackage{marginnote}
\usepackage{amssymb}
\usepackage{epstopdf}
\usepackage{color}
\usepackage{amsthm}
\usepackage{caption}
\graphicspath{{./fig_pdf_reduced/}}

\begin{document}
\title{Free-breathing and ungated dynamic MRI using navigator-less spiral SToRM}

\author{Abdul Haseeb~Ahmed, 
        Ruixi~Zhou,
        Yang~Yang,
        Prashant~Nagpal,
        Michael~Salerno,
        and~Mathews~Jacob 
\thanks{This work is supported by NIH under Grants R01EB019961, R01HL131919-01A1.}
\thanks{Abdul Haseeb Ahmed is with the Department
of Electrical and Computer Engineering, University of Iowa, Iowa City, USA. 
e-mail:abdul-ahmed@uiowa.edu}
\thanks{Ruixi Zhou, and Michael Salerno are with the Department of Biomedical Engineering, University of Virginia, Charlottesville, USA.
emails:rz3hu@virginia.edu and MS5PC@hscmail.mcc.virginia.edu}
\thanks{Yang Yang is with the Institute and Department of Radiology, Icahn School of Medicine at Mount Sinai, New York, USA. 
e-mail:yy5cc@virginia.edu}
\thanks{Prashant Nagpal is with the Department
	of Radiology, University of Iowa, Iowa City, USA. 
	e-mail:prashant-nagpal@uiowa.edu}
\thanks{Mathews Jacob is with the Department
	of Electrical and Computer Engineering, University of Iowa, Iowa City, USA. 
	e-mail:mathews-jacob@uiowa.edu}
\thanks{Manuscript is accepted in IEEE Transactions on Medical Imaging.}
}

\maketitle
\begin{abstract}
We introduce a kernel low-rank algorithm to recover free-breathing and ungated dynamic MRI from spiral acquisitions without explicit k-space navigators. It is often challenging for low-rank methods to recover free-breathing and ungated images from undersampled measurements; extensive cardiac and respiratory motion often results in the Casorati matrix not being sufficiently low-rank. Therefore, we exploit the non-linear structure of the dynamic data, which gives the low-rank kernel matrix. Unlike prior work that rely on navigators to estimate the manifold structure, we propose a kernel low-rank matrix completion method to directly fill in the missing k-space data from variable density spiral acquisitions. We validate the proposed scheme using simulated data and in-vivo data. Our results show that the proposed scheme provides improved reconstructions compared to the classical methods such as low-rank and XD-GRASP. The comparison with breath-held cine data shows that the quantitative metrics agree, whereas the image quality is marginally lower. 
\end{abstract}

%
\IEEEpeerreviewmaketitle
\section{Introduction}
\par Breath-held cine MRI is an integral part of clinical cardiac exams. It is widely used for the anatomical and functional assessment of the heart.  Diagnostic cine images require breath holding which results in a long scan time to achieve better spatial and temporal resolution. It is often challenging for children, patients with heart failure and patients with respiratory complications such as chronic obstructive pulmonary disease (COPD) \cite{copd}.  In addition, multiple breath holds along with intermittent pauses also prolong the scan time, adversely impacting patient comfort and compliance. The scans from different slices may also suffer from inconsistencies between breath-held positions {\cite{inconsist_bH}. The acceleration of breath-held cine MRI has been the subject of extensive research in the recent past.} Classical approaches include parallel MRI, where the diversity of coil sensitives are exploited to reduce the breath-held duration.{\textcolor{black}{Recent approaches further improve the performance by exploiting the structure of x-f space \cite{blast}, sparsity \cite{sparse},
low-rank property \cite{zhao2012,ktslr}, low-rank +sparsity \cite{otazo}, learned dictionaries \cite{bcs}, motion-compensated methods \cite{master2013}, deep learning methods \cite{schlem} and kernel low-rank methods \cite{nakarmi2017kernel}}. When the subjects cannot hold their breath, a standard alternative is real-time imaging, which does not require breath holding or ECG gating. However, these methods have been shown to sacrifice spatial and/or temporal resolution \cite{uribe2007whole,feng2016non}. Another approach is the use of diaphragmatic navigators, which restricts the acquisition to images in the specific respiratory phase \cite{peters20082d}.  The drawbacks of these schemes include respiratory gating efficiency and variability in the scan time. Several methods that rely on radial acquisitions were introduced in recent years to estimate the cardiac and respiratory phases from the central k-space regions using band-pass filtering \cite{xdgrasp}. These methods usually require careful selection of receiver coils to obtain self-gating signals, as each coil has different sensitivity to cardiac and respiratory motions. The data is then binned to the respective phases, followed by reconstruction using compressed sensing \cite{xdgrasp} or low-rank tensor methods \cite{christodolou}. Methods that rely on respiratory motion compensation followed by binning have also been introduced to improve computational efficiency \cite{zhou2018}.  A challenge with these approaches is  the dependency on  the phase estimation using band-pass filtering that relies on cardiac and respiratory rates, which may degrade in the presence of irregular respiratory motion or arrhythmia {\cite{sunritatci,recon_arryth}}. Since these methods rely on the explicit segmentation of the data into their respective phases, the applicability of  these schemes for arrhythmia \cite{Chitiboi18} or for non-cardiac applications (e.g, speech) is not straightforward. 

We recently introduced the smoothness regularization on manifold (SToRM) approach, which enables ungated cardiac cine imaging in the free-breathing mode using  radial acquisitions \cite{storm,sunritatci}. SToRM algorithm assumes that the images lie on a smooth and low-dimensional manifold, parameterized by a few variables (e.g. cardiac and respiratory phases). We note that the smooth manifold/surface model is a non-linear generalization of the linear subspace/low-rank models. These models represent the dynamic dataset more efficiently as compared to the subspace models, which result in reduced blurring in free-breathing applications with extensive cardiac and respiratory motion \cite{storm,sunritatci}. The manifold prior facilitates the implicit sharing of data between images in the dataset that have similar cardiac or respiratory phases, which is an alternative to explicit motion-resolved strategies \cite{xdgrasp,christodolou}. 
While this approach does not perform  explicit binning of data as in other studies \cite{xdgrasp,christodolou}, it still exploits the similarity of images in the time series and can be viewed as a soft-binning strategy; a particular image is not assigned to any phase, but the inter-frame weights indicate the similarity of the image with other images in the time series.
Since the framework does not require complex processing steps that assume the periodicity of the cardiac/respiratory motion, it is readily applicable to several dynamic applications, including speech imaging, as shown in previous work \cite{storm}, and cardiac applications involving arrhythmia. We  note that  there are similar manifold regularization schemes that also rely on the non-linear structure of the data to recover cardiac MRI data \cite{nakarmi2017kernel,bilinear}. The main difference of our algorithm from past kernel low-rank methods \cite{nakarmi2017kernel} is that we do not require an explicit evaluation of the image feature maps. The work \cite{bilinear} relies on a sparse optimization scheme to recover the Laplacian matrix from navigator data,  which is used to recover the data as in our work \cite{sunritatci}. 

Our previous implementation, which we refer to as SToRM:Self-Nav, as well as \cite{nakarmi2017kernel,bilinear} relied on explicit radial k-space navigators to estimate the manifold structure. Specifically, a few radial spokes with the same orientations are played out periodically. The manifold Laplacian estimated from the navigator data is used to recover the images \cite{sunritatci}.  Compared to the short radial readouts, the longer spiral readouts considered in this work offers improved sampling efficiency; this approach enables the acquisition of more k-space samples in a given scan time. These longer readouts along with higher flip angles also offer improved myocardial contrast. However, the direct use of SToRM:Self-Nav in our setting results in a large overhead; this approach would require the acquisition of one navigator readout for every frame (corresponding to 3-4 spiral interleaves). In addition, the navigated approach cannot be readily applied to golden angle sequences implemented on several scanners without dedicated navigators. To minimize the above problems, we generalize the SToRM algorithm to recover free-breathing and ungated cardiac MRI data from a variable-density spiral gradient echo (GRE) acquisition without any navigators. Since the navigators are not available, we use the kernel low-rank penalty in the matrix completion setting to fill the missing entries. To improve computational efficiency, we rely on a two-step approach, where the low-rank matrix completion is first applied to low-resolution data. This step is computationally efficient because the size of the images are small. Once the low-resolution data is obtained, the manifold Laplacian estimated from this data is used to recover the high-resolution images. We rigorously validate the spiral SToRM approach against conventional algorithms as well as breath-held cine, both quantitatively and qualitatively using simulated as well in-vivo multi-slice data.
\begin{figure*}
	\center
	\includegraphics[width=0.75\textwidth]{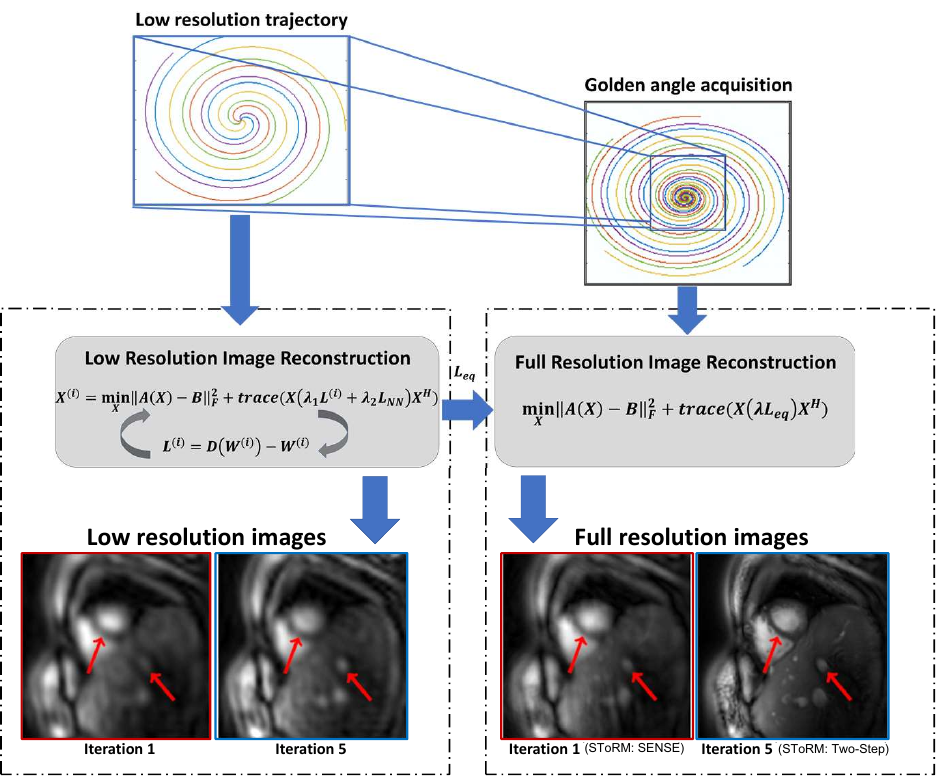}\\  
	\caption{\small \color{black} Outline of the SToRM: Two-Step method. Free-breathing and ungated data is acquired using golden angle interleaved spiral trajectories. We rely on a two-step strategy, where a low-resolution dataset is first recovered from the central k-space regions denoted by the blue box. Since this region is still not fully sampled, kernel low-rank regularization is used to recover the images. As described in the text, this iterative strategy yields the Laplacian matrix as a by-product. Once the Laplacian is available, the high-resolution dataset is estimated from all of the k-space samples by solving Eq. \eqref{imageupdate}. The first image with red border in the left panel corresponds to the low-resolution image recovered by the first iteration of the kernel low-rank algorithm, which corresponds to a SToRM:SENSE method. The Laplacian matrix estimated from this result (iteration 1 with red border) is used to recover the high-resolution data, indicated by the first image with a red border in the right panel. By contrast, iterating the kernel low-rank algorithm provides more details, as shown by the second image in the left panel with a blue border. The recovery using the Laplacian from this estimate, termed as SToRM:Two-Step, yields improved image quality, as shown in the second image in the right panel with a blue border.}
	\label{fig1}
\end{figure*}

\vspace{-0.1cm}
\section{Background}

The SToRM framework relies on the manifold structure of images in the real-time cardiac MRI. The main focus of this work is to extend the SToRM framework with explicit k-space self-gating navigators to a navigator-free setting, which increases the sampling efficiency. The proposed navigator-free SToRM algorithm is enabled by the variable density sampling offered by the spiral gradient echo (GRE) acquisition. The GRE acquisition is free from banding artifacts and does not require additional frequency scouts, which are needed to minimize banding artifacts in steady state free precession (SSFP) sequences on 3T scanners. In comparison to the navigated radial acquisition scheme in earlier work \cite{storm}, the spiral acquisition scheme offers higher sampling efficiency and signal-to-noise ratio. In addition, we propose to exploit the central k-space regions that are densely sampled relative to the radial settings, which would eliminate the need for navigator-based acquisition to determine the Laplacian matrix. We now briefly describe the SToRM framework.
\vspace{-0.2cm}  
\subsection{Overview of SToRM framework \cite{storm}}
The SToRM scheme models the images in the dataset as points on a low-dimensional smooth manifold in high-dimensional space, which is equal to the number of pixels in each image. The SToRM algorithm relies on minimizing a smoothness prior on the manifold  of images $\mathbf x_i$ in the time series to exploit this structure: 
\begin{equation}
\label{unidimensional}
\int_{\mathcal M} \|\nabla x\|^2 \approx \frac{1}{2}\;\sum_{i,j=1}^k \mathbf w_{i,j}\; \|\mathbf x_i -\mathbf x_j\|^2 = {\textrm{ trace}}(\mathbf X\,\mathbf L\, \mathbf X^H).
\end{equation}
Here, $\mathbf X$ is the Casorati matrix, whose columns correspond to $\mathbf x_i; i=1,..,k$. The weights $ \mathbf w_{i,j}$ specify the neighborhood structure on the points/images, or equivalently the similarity between images in the dataset. 

In our previous work, we relied on 4 radial navigator spokes with the same orientation that were played out periodically (repeated every 10 spokes) to estimate the weights $\mathbf w_{i,j}$ using kernel low-rank estimation \cite{storm,sunritatci}. These readouts were referred as k-space navigators. The rest of the radial spokes are played out in the golden angle view ordering. Denoting k-space data from the navigator spokes at the $i^{\rm th}$ image by $\mathbf z_i $, the weights are estimated from the equation as
\begin{equation}
\label{Wcomp}
\mathbf w_{ij} =\mathbf e^{-\frac{\|\mathbf z_{i}-\mathbf z_j\|^2}{\sigma^{2}}}.
\end{equation}
Note that the above choice assigns higher weights to image pairs $\mathbf x_i$ and  $\mathbf x_j$, if the differences of their k-space navigators specified by $\|\mathbf z_i-\mathbf z_j\|^2$ indicate that they are neighbors on the manifold. Here,  $\sigma$ is a parameter that controls the smoothness of the manifold.  $\mathbf L = \mathbf D - \mathbf W$ is the Laplacian matrix in \eqref{unidimensional}. Here, $\mathbf D$ is a diagonal matrix with elements defined as $\mathbf D_{ii} = \sum_j \mathbf W_{ij}$.  

Once $\mathbf L$ is available, SToRM performs the joint recovery of the images in the dataset by solving the following problem:
\begin{eqnarray}
\mathbf X^{*} =\arg \min_{\mathbf X} \|\mathcal A(\mathbf X)-\mathbf B\|^{2}_{F} + \lambda~ \mathrm{trace}(\mathbf X {\mathbf L} \mathbf X^{H}).
\label{l2problem}
\end{eqnarray}
Here $\mathcal A$ is the measurement operator that accounts for the multichannel spiral sampling of the columns of $\mathbf X$, which are the image frames.

\vspace{-0.2cm}
\subsection{Bandlimited SToRM model}

We consider a bandlimited surface model, where the images $\mathbf x_1, ..\mathbf x_N$ are modeled as high-dimensional points on a smooth surface \cite{levelsetosher,sunritatci}. We model the surface as the zero level-set of a band-limited function  $\psi(\mathbf x)$:
\begin{equation}\label{key}
\mathcal S = \{\mathbf x| \psi(\mathbf x) = 0\},
\end{equation}
where $\psi(x)$ are linear combination of exponentials, whose frequencies are supported at $\mathbf k_1, \ldots, \mathbf k_P$ on a discrete lattice. 

With the bandlimited assumption, we have shown that the exponential feature maps $\Phi(\mathbf x)$ of the images specified by 
\begin{equation}
\label{expmaps}
\phi(\mathbf x) = \begin{bmatrix} \frac{1}{\sigma^2\|\mathbf k\|}\exp\left(j \mathbf k_1^T \mathbf x\right)\\\vdots \\
\frac{1}{\sigma^2\|\mathbf k\|}\exp\left(j \mathbf k_P^T \mathbf x\right),
\end{bmatrix}
\end{equation}
live in a low-dimensional subspace. This implies that the feature matrix  
\begin{eqnarray}
\Phi(\mathbf X) = \begin{bmatrix} \phi(\mathbf x_1) & \phi(\mathbf x_2)& \ldots &  \phi(\mathbf x_N).
\end{bmatrix}
\end{eqnarray}
is low-rank \cite{sunritatci}. If the rank is $r$, we can find an orthonormal basis $\mathbf Q \in \mathbb C^{(n-r)\times n}$ of the null-space such that $\Phi(\mathbf X)\mathbf Q =0$. 

Similar to PSF methods \cite{zhao2012}, we have estimated the above subspace from the inverse Fourier transform of the navigator readouts, denoted by $\mathbf Z$. Note that in the radial setting $\mathbf Z$ corresponds to projections of the $\mathbf X$ along specific orientations. We assume that the null-space $\mathbf Q$ of the feature matrix $\Phi(\mathbf X)$ to be the same as the null-space of $\Phi(\mathbf X)$. We estimate $\mathbf Q$ by picking the $n-r$ lowest singular vectors of the kernel matrix $\mathcal K(\mathbf Z)= \Phi(\mathbf Z)^H\Phi(\mathbf Z)$.

\section{Proposed approach}
The above approach works well with radial k-space navigator lines, which are fully sampled along the readout direction. In addition 2-4 lines are often needed to reliably estimate the null-space, which reduces the sampling efficiency. In this work, we propose to use a spiral trajectory which provides improved sampling efficiency than radial acquisition. We also eliminate the need for k-space navigators to further improve the sampling efficiency. 
\vspace{-0.2cm}
\subsection{\color{black}Kernel low-rank matrix completion for spiral cine data}
\label{lapest}	

We propose to recover the images $\mathbf x_1, ..\mathbf x_N$ from their undersampled measurements by relying on kernel low-rank matrix completion: 

\begin{equation}
\begin{split}
\label{lapform}
\mathbf X^{*} =\arg \min_{\mathbf X} \|\mathcal A(\mathbf X)-\mathbf B\|^{2}_{F} + \lambda_{1}~\left\|\Phi(\mathbf X)\right\|_*.
\end{split}
\end{equation}

Due to extensive cardiac and respiratory motion, the matrix $\mathbf X$ may have a higher rank since the images may not lie on a subspace with small dimension. However, the images may lie on a smooth surface, resulting in the feature matrix being low-rank. The second term in \eqref{lapform} is the nuclear norm of the non-linear features of the images $\mathbf x_i$, which promotes the low-rank nature of $\Phi(\mathbf X)$. This prior forces the feature maps $\Phi(\mathbf x_i)$ to a subspace, which is equivalent to encouraging the images $\mathbf x$ to lie on smooth surface specified by $S$. This approach is a non-linear generalization to classical low-rank/subspace models \cite{ktslr}, which are widely used in dynamic imaging. While this approach was used in the denoising setting \cite{sunritatci}, the utility of this scheme in completing dynamic MRI datasets has not been reported. Results are available for polynomial varieties \cite{gregvariety}, but their utility in medical imaging have not been explored.
\vspace{-0.2cm}
\subsection{Iterative reweighted algorithm for matrix completion}
\label{irls}
The direct implementation of eq. \eqref{lapform} would require the non-linear mapping between the images $\mathbf x_i$ and their features $\phi(\mathbf x_i)$, as well as their inverse  \cite{nakarmi2017kernel}. However, this approach is computationally infeasible in our setting since the dimension of the feature matrix $\Phi(\mathbf X)$ is too large. Therefore, we use an algorithm that relies on the Gram matrix of $\Phi(\mathbf X)$, denoted by $\mathcal K(\mathbf X)=\Phi(\mathbf X)^H\Phi(\mathbf X)$, which is referred as the kernel matrix.  For the specific choice of exponential maps as in \eqref{expmaps}, the entries of $\mathcal K(\mathbf X)$ can be computed directly as:
\begin{equation}
\label{kernel}
\left[\mathcal K(\mathbf X) \right]_{i,j}= \exp\left(- \frac{\|\mathbf x_i - \mathbf x_j\|^2}{2\sigma^2}
\right).
\end{equation}
without requiring the evaluation of the features $\Phi(\mathbf x_i); i=1,..,N$. This approach is widely known as the kernel trick in machine learning \cite{kernelbook}. Specifically, we use the iterative reweighted least squares algorithm with gradient linearization \cite{sunritatci} to obtain an alternating algorithm to solve \eqref{lapform}. This algorithm alternates between
\begin{equation}
\begin{split}
\label{imageupdate}
\mathbf X^{(n)} =\arg \min_{\mathbf X} \|\mathcal A(\mathbf X)-\mathbf B\|^{2}_{F} +  \lambda_{1}\textrm{ trace}(\mathbf X ~\mathbf L^{(n)}\mathbf X^{H})
\end{split}
\end{equation}
and update of the matrix $\mathbf L^{(n)}$:
\begin{equation}
\label{lapupdate}
\mathbf L^{(n)} = \mathbf D^{(n)} - \mathbf W^{(n)}.
\end{equation}
Here, the weight matrix at the $n^{\textrm {th}}$ iteration is specified by
\begin{equation}
\label{wtupdate}
\mathbf W^{(n)} = -\frac{1}{\sigma^2}\mathcal K\left(\mathbf X^{(n-1)} \right) \odot \left[\mathcal K\left(\mathbf X^{(n-1)}\right) + \gamma \mathbf I\right]^{-\frac{1}{2}},
\end{equation}
and $\mathbf D^{(n)}$ is the diagonal matrix with diagonal entries $\mathbf D^{(n)}_{ii} = \sum_j \mathbf W^{(n)}_{i,j}$.
\\
Note that the above approach aims to recover a dataset with approximately 400 images. Hence, the computation cost associated with this scheme is high, especially in the high spatial resolution settings. In later section, we have compared our approach with full resolution Laplacian estimation approach to show the benefit of our scheme. 

\subsection{{\textcolor{black}{Improve computational efficiency using SToRM: Two-Step}}}
\label{twostep}
We rely on a two step approach to keep the computational complexity of the algorithm minimal. We call this approach as SToRM: Two-Step. First we propose to recover the low-resolution images from the central k-space samples as shown in  Fig. \ref{fig1}. Note that the low-resolution images are still undersampled. We use the formulation in \eqref{lapform} with the addition of a Tikhnonov temporal prior:
\begin{equation}
\begin{split}
\label{lapnew}
\mathbf X_L^{*} =\arg \min_{\mathbf X_L} \|\mathcal A_L(\mathbf X_L)-\mathbf B_L\|^{2}_{F} + \lambda_{1}~\left\|\Phi(\mathbf X_L)\right\|_*  + \\ \lambda_{2}~ \underbrace{\sum_i \|\mathbf x_{(i+1)}-\mathbf x_i\|^2}_{\mathrm{{\textrm{trace}}}(\mathbf X_L ~{\mathbf L_{\textrm{tik}}} ~\mathbf X_L^{H})}.
\end{split}
\end{equation}
Here, $\mathcal A_L$ and $\mathbf B_L$ are the forward model and the measured multichannel k-space data corresponding the the central k-space regions. By exploiting the similarity of adjacent temporal neighbors, this approach is expected to further improve performance over \eqref{lapform}. The smaller size of the images translates to a faster algorithm.{\color{black} Note that the Tikhonov prior can be rewritten as $\mathrm{{\rm trace}}(\mathbf X_L {\mathbf L_{\rm {tik}}} \mathbf X_L^{H})$, where $\mathbf L_{\rm{tik}}$ is the matrix with block diagonal matrix with entries as $[1,-2,1]$.} Once the above algorithm converges, the estimated Laplacian matrix is then used to recover the high-resolution image frames from their undersampled measurements by solving \eqref{imageupdate} as shown in Fig. \ref{fig1}. The iterative algorithm for solving this cost function is similar to the one in Section \ref{irls}, where the step \eqref{imageupdate} is modified as 
{\color{black}{
\begin{equation}
\begin{split}
\label{imageupdatenew}
\mathbf X_L^{(n)} =\arg \min_{\mathbf X_L} \|\mathcal A_L(\mathbf X_L)-\mathbf B_L\|^{2}_{F} + {\textrm{ trace}}(\mathbf X_L~\mathbf L_{\textrm {eq}}^{(n)}~\mathbf X_L^{H}),
\end{split}
\end{equation}
}}
where $\mathbf L_{\textrm {eq}}^{(n)} = \lambda_{1}~\mathrm{\mathbf L^{(n)}}+\lambda_{2}~\mathrm{\mathbf L_{\rm {tik}}}$. Since we do not alternate between the Laplacian update and the image update in the high-resolution setting, we obtain a fast algorithm. Once the above iterative algorithm converges, we use the $\mathbf L_{\rm eq}$ matrix to recover the high-resolution data by solving the quadratic optimization scheme specified by \eqref{imageupdate}.  This two-step approach is illustrated in Fig. \ref{fig1}. 

\section{Experimental details}
\subsection{Datasets}

We use the following datasets for the experimental evaluation of the proposed algorithm:

\noindent \underline{Simulated Dataset: } A  retrospective ECG-gated, breath-held cardiac MRI is used to create simulated ungated, free-breathing data, as described in by Zhao et al.\cite{zhao2012}. The ground truth breath-held SSFP dataset is warped in space and time to mimic respiratory motion and temporally varying heart rate. Please see Zhao et al. \cite{zhao2012} for details. The deformed datasets are combined to form an image sequence with multiple cardiac cycles. This free-breathing dataset has a reasonable amount of inter-frame motion due to respiratory dynamics. The dataset has 200 phase encodings, 256 samples per readout, and 256 temporal frames. FOV= $273$mm $\times 350$mm, spatial resolution$=1.36$mm $\times 1.36$mm and TR$=3$ms. For simulated data, variable density spirals are used with 12 spirals per frame. This simulated dataset enables the quantitative comparison of methods, especially in the free-breathing setting where ground truth is not available.

\noindent\underline{In-vivo Datasets: } Six single slice cardiac data and five whole-heart multi-slice datasets were collected in the free-breathing mode using a golden angle spiral trajectory. We compare the proposed scheme with breath-held whole heart bSSFP Cartesian acquisition on five datasets with 10-13 slices to cover the whole heart. Acquisition parameters: TR/TE= 3.1ms-3.4ms/1.18ms-1.28ms, flip angle= 31-39 degrees. All acquisitions except one were performed on a 3T scanner (MAGNETOM Prisma, Siemens Healthineers, Erlangen, Germany). One dataset was acquired on the GE 3T scanner. Image datasets were acquired using the standard body phased-array RF coil. Subjects included three females (age: 25-27) and eight males (age: 20-30) with short-axis view cine data. The institutional review board at the local institution approved all the in-vivo acquisitions, and written consent was obtained from all subjects. The sequence parameters were: TR/TE= 7.8 ms/1 ms, FOV= 320 mm, Base resolution= 256, Bandwidth= 390 Hz/pixel, flip angle= 15 degrees, slice thickness= 8 mm. Dual-density spirals were generated using a Fermi function with a k-space density of 0.2x Nyquist for the first 20\% of the trajectory and an ending density of 0.02x Nyquist \cite{zhou2019free}. The spirals were continuously acquired with rotation of the trajectory by the golden angle between spirals. Off-resonance effects were minimized by using a short spiral readout duration (5 ms) and by using the vendor-provided cardiac shim routine. Post-acquisition, five spirals per frame were binned to obtain the temporal resolution of 40 ms. For multi-slice data, we have collected 10-13 slices from apex to base to cover the whole heart.
The parameters of our reconstruction algorithm were manually optimized on one dataset and kept fixed for rest of the datasets: $\lambda_{1}$=0.01, $\lambda_{2}$=1e-5, $\sigma=4.5$, and $\lambda$=0.025. $\lambda_{1}$ and $\lambda_2$ involve the trade off between blurring and aliasing artifacts. Here, $\sigma$ is the width of the temporal kernel. We notice that the results are not too sensitive to $\sigma$; kernel low-rank regularization makes the algorithm relatively insensitive to kernel width.
The above mentioned, $\gamma$ $(=100)$ is a smoothing parameter that is decreasing in each iteration. In order to have a long-run stability of the algorithm, we decrease  until it approaches to pre-determined minimum value. Further details are mentioned in Ongie et.al. \cite{ongie2017} 
\begin{figure}
	\centering
	\includegraphics[ scale=0.5]{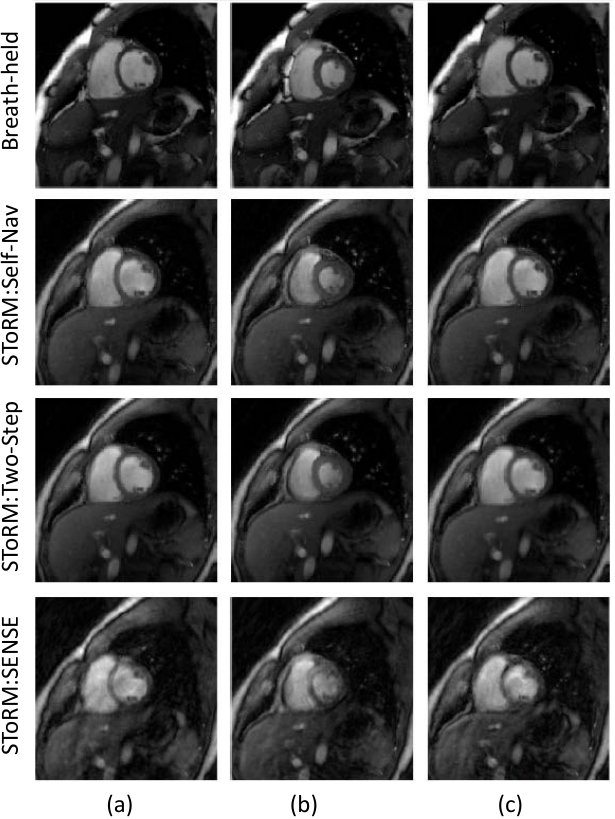}\\   
	\caption {\small {Visual comparison of SToRM:Two-Step with breath-held SSFP CINE and SToRM:Self-Nav reconstructions. Breath-held data was acquired in the end of inspiration. SToRM: Two-Step gives comparable image quality to SToRM:Self-Nav and breath-held results. However, we observe better sharpness in the breath-held results as compared to the SToRM: Two-Step. Whereas, aliasing artifacts are observed in SToRM:SENSE results.}}
	\label{fig2}
\end{figure}

\begin{figure}[h]
	\centering
	\vspace{0cm}
	\includegraphics[width=0.4\textwidth]{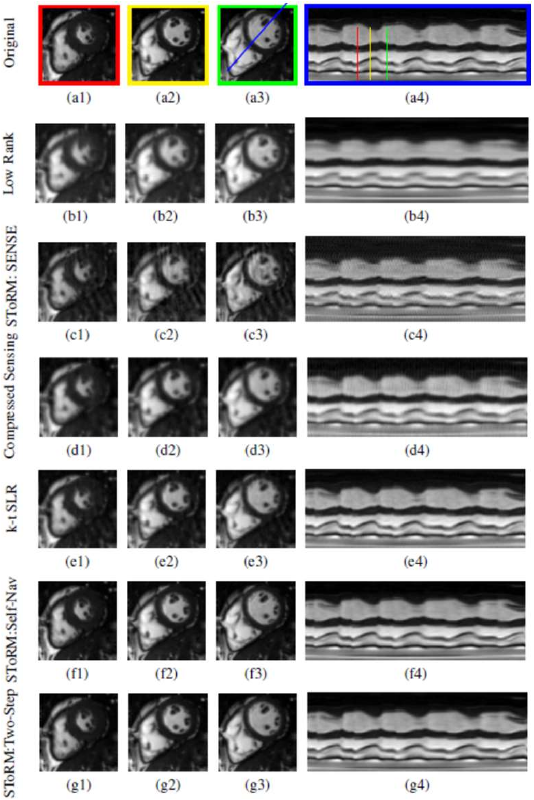}\\
	\caption{\small \color{black} Performance of the algorithm using  a simulated short-axis cardiac cine dataset. We compare the low-rank algorithm (b1-b3), SToRM: SENSE method (c1-c3), the compressed sensing method (d1-d3), k-t SLR (e1-e3), SToRM: Self-Nav (f1-f3) and the proposed method (g1-g3). Each scheme (k-t or non-binning method) is compared against the original dataset (a1-a3). This dynamic dataset is retrospectively undersampled using a golden angle spiral sampling pattern. Three cardiac phases are picked from each reconstruction method and correspond to end of systolic, mid phase, and end of diastolic, as shown by red, yellow, and green lines in the time profile (a4).  The time profiles in the last column are shown for the entire time series, along the line passing through the left ventricle and right ventricle shown in (a3). We observe that the proposed method provides reconstructions with lower spatial and temporal blurring compared to low-rank, SToRM:SENSE, k-t SLR and compressed sensing methods. It gives comparable image quality to SToRM:Self-Nav. Table \ref{table1} shows a quantitative comparison of the methods using SER, HFEN, SSIM, and GPC metrics computed around the cardiac region.Since XD-GRASP uses different reconstruction strategy,so we have done separate comparison with XD-GRASP method.}
	\label{fig3}
\end{figure}
\begin{table*}[h]
	\vspace{0cm}
	\centering
	\hspace{-0.5cm}
	\caption{\small \color{black} Quantitative comparison of the methods on simulated data in Fig. \ref{fig3} using the signal-to-error ratio (SER), normalized high frequency error (HFEN), structural similarity index (SSIM) and global phase coherence (GPC) metrics. All of these metrics are computed in a square region of interest around the cardiac region. Higher values  of the above-mentioned performance metrics correspond to better reconstruction except for the HFEN, where a lower value is better. These comparisons show that the proposed scheme performs better than the other methods except for SToRM: Self-Nav, where dedicated k-space navigators are used to estimate the Laplacian matrix.}
	{\color{black}
	\label{table1}

	\begin{tabular}{|c|c|c|c|c|}
		\hline
		Method & SER & SSIM & HFEN  & GPC\\ \hline
		Low Rank  & ${16.60 \pm 0.90}$    & ${0.80 \pm 0.03}$          & ${0.42 \pm 0.05}$   & ${115 \pm 33}$\\ \hline
		SToRM: SENSE&  ${16.47 \pm 0.86}$    & ${0.71 \pm 0.03}$          & ${0.47 \pm 0.04}$   &  ${112 \pm 60}$\\ \hline
		Compressed Sensing&  ${17.90 \pm 1.60}$    & ${0.80 \pm 0.06}$          & ${0.38 \pm 0.07}$   &  ${106 \pm 29}$\\ \hline
		kt-SLR&  ${18.60 \pm 0.70}$    & ${0.87 \pm 0.02}$          & ${0.25 \pm 0.01}$   &  ${204 \pm 35}$\\ \hline
		SToRM: Self-Nav&  ${25.01 \pm 1.41}$    & ${0.94 \pm 0.02}$      & ${0.128 \pm 0.02}$   & ${428 \pm 113}$\\ \hline
		SToRM: Two-Step & \boldmath${25.77 \pm 1.50}$    & \boldmath${0.95 \pm 0.02}$          & \boldmath${0.107 \pm 0.02}$    &\boldmath ${475 \pm 110}$\\ \hline
	\end{tabular}
	}
	\vspace{0cm}
\end{table*}

\subsection{Imaging Experiments}
All the results were generated using a single node of a high-performance Argon Cluster at the University of Iowa, equipped with an Intel Xeon CPU with 28 Cores at 2.40 GHz with 128 GB of memory running on Red Hat Linux MATLAB R2016b. The reconstruction time of the proposed method was between 8 to 10 minutes (400 time frames). This reconstruction is for single slice reconstruction. Low resolution stage takes around 65\% of the time and rest is spent in the final reconstruction, as shown in the Table. \ref{table3}.

\noindent \underline{Coil selection and compression:} We acquired the dataset using 34 coils. However, we excluded the coils with low sensitivities in the region/slice of interest. We used an automatic algorithm to pre-select the 10 best coil images that provided the best signal to noise ratio in the heart region; we observed that removing the unreliable coils resulted in improved reconstructions \cite{zhou2019free}. This algorithm binned the k-space data from several images to recover the low-resolution coil images. We then used PCA-based coil combination using SVD such that the approximation error was $<5\%$. In most cases, we noted that 5-6 coils were sufficient to bring the approximation error to $<5\%$. The coil sensitivity maps were estimated from these coil-combined virtual channels using the method designed by Walsh et al. \cite{walsh2000adaptive} and assumed to be constant over time. Our experiments (not included in the paper) show that this coil combination has minimal impact on image quality. The main motivation for the combination was to reduce the memory requirement so that it fit on our GPU device, which significantly reduced the computational complexity.

\noindent\underline{Performance Metrics:} We used four quantitative metrics to compare our method against the existing schemes: 
\begin{itemize}
	\item Signal-to-Error Ratio (SER):
	\begin{equation}
	\mathbf{SER} = 20\log_{10} {\frac{||\mathbf{x}_{\textrm{ orig}}||_{2}}{||\mathbf{x}_{\textrm {orig}}-\mathbf{x}_{\textrm{rec}}||_{2}}},
	\end{equation}
	where $||\cdot||_{2}$ donates the $\ell_{2}$ norm, and $\mathbf{x}_{orig}$  and $\mathbf{x}_{rec}$ denote the original and the reconstructed images, respectively.
	\item Normalized High Frequency Error (HFEN) \cite{ravishankar2011}: This measures the quality of fine features, edges, and spatial blurring in the images and is defined as: {\color{black}{
	\begin{equation}
	\mathbf{HFEN} = 20\log_{10} {\frac{||\textrm{LoG}(\mathbf{x}_{\textrm {orig}})-\textrm{LoG}(\mathbf{x}_{\textrm {rec} })||_{2}}{||\textrm{LoG}(\mathbf{x}_{\textrm {orig}})||_{2}} },
	\end{equation}
	}}
	where $\textrm{LoG}$ is a Laplacian of Gaussian filter that captures edges. We use the same filter specifications as in Ravishankar et al. \cite{ravishankar2011}: kernel size of 15 $\times$ 15 pixels, with a standard deviation of 1.5.
	\item The Structural Similarity index (\textbf{SSIM}) is a perceptual metric introduced by Wang et al. \cite{wang2004image}. We used the toolbox introduced by Wang et al.  \cite{wang2004image}: with default contrast values, Gaussian kernel size of 11 $\times$ 11 pixels with a standard deviation of 1.5 pixels.
	\item Global phase coherence (\textbf{GPC}) index \cite{bmr08} provides a measure of image sharpness by estimating the volume of all possible phase functions associated with the measured modulus, which produces images that are not less likely than the original image. The likelihood is measured with the total variation implicit prior, and is numerically evaluated using a Monte-Carlo simulation. We used the toolbox introduced by Blanchet et al. \cite{bmr08} to compute this index for our images.
\end{itemize}
Higher values  of the above-mentioned performance metrics correspond to better reconstruction, except for the HFEN, where a lower value is better.
\subsection{Algorithms for comparisons}

We have used both simulated and in-vivo data to compare the following algorithms:
\begin{itemize}
	\item SToRM: Two-Step (Proposed): The manifold Laplacian is estimated iteratively by alternating between the estimation of the Laplacian matrix and the update of the images on the low-resolution data. Once the Laplacian is obtained, the high-resolution images are recovered by solving \eqref{imageupdate} using all the k-space samples.
	
	\item SToRM: Single-step (Full resolution): we alternate between \eqref{imageupdate} and \eqref{lapupdate} on  the high-resolution data. This approach may offer improved quality than SToRM:Two-Step since the high-resolution details can potentially yield improved estimation of the Laplacian, and consequently improved results. However, this approach is associated with higher computational complexity.
	
	\item SToRM:Original: We estimate the Laplacian matrix from ground truth data. This approach is  only possible in the simulated setting, and provides an upper bound for the image quality. 
	
	\item SToRM: SENSE: In this method, we  estimate the Laplacian matrix from the CG-SENSE reconstructions. Equation \eqref{lapnew} is modified as: 
			\begin{equation}
			\begin{split}
			\label{low_CGSENSE}
			\mathbf X_S^{*} =\arg \min_{\mathbf X_S} \|\mathcal A_S(\mathbf X_S)-\mathbf B_S\|^{2}_{F} + \|\mathbf X_S\|^{2}_{F},
			\end{split}
			\end{equation}
	The manifold Laplacian is recovered from $\mathbf X_S$. This Laplacian matrix is then used to recover the high-resolution images by solving \eqref{imageupdate} from all the k-space samples.

	\item SToRM: Self-Nav \cite{sunritatci}: The manifold Laplacian is recovered from the self-gating navigators acquired in k-space, followed by \eqref{imageupdate} using all the k-space samples.
	\item XD-GRASP \cite{xdgrasp}:  This self-gated strategy estimates the cardiac and respiratory phases from the center sample of k-space regions \cite{xdgrasp}. It estimates the cardiac and respiratory signals by filtering the central regions with different band-pass filters, each corresponding to the cardiac and respiration frequencies.  We used the author-provided MATLAB code for XD-GRASP implementation \cite{xdgrasp}.
	\color{black}{
	\item k-t SLR \cite{ktslr}: The image time series is recovered by Schatten p-norm (p=0.5) and total variation regularization minimization. We have tuned the sparsity and low rank regularization
	parameters to get the optimum results on our dataset. 
	\item Low-Rank \cite{zhao2012,ktslr}: The image time series is recovered by nuclear norm minimization. The nuclear norm minimization approach models the images as points living on a subspace and we are setting the sparsity regularization parameter to zero.}

	\item Compressed sensing \cite{sparse}: The image time series is recovered by $l_1$ sparsity regularization with total variation.	
	
\end{itemize}

For image quality and ejection fraction comparison, all data sets were assessed by a board certified radiologist and a cardiovascular imager in a blinded manner. Image quality was evaluated on a 5-point score ranging from 1 (poor and not acceptable clinically) to 5 (excellent clinically). The ejection fraction comparison between breath-held and SToRM: Two-Step results, was performed using a 2-way analysis of variance (ANOVA) analysis. Statistical analysis was performed using SAS software (version $9.4$; SAS Institute Inc., Cary, NC).

\begin{figure}[htb]
	\center
	\vspace{0cm}
	\includegraphics[width=0.35\textwidth]{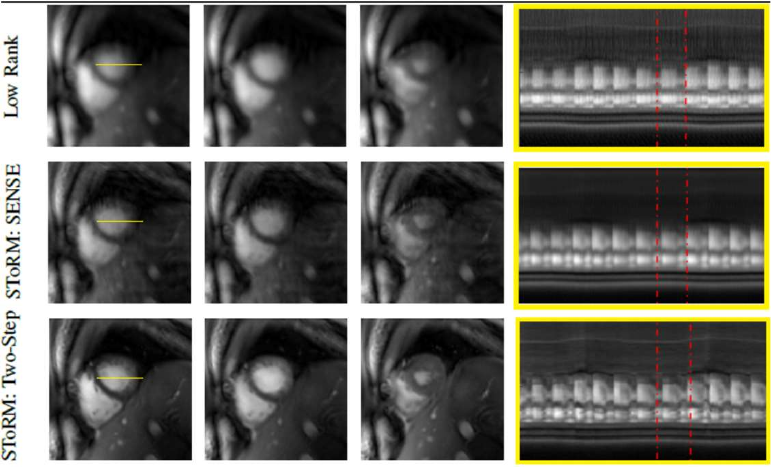}\\
	\caption {\small \color{black} Comparison against free-breathing methods that do not use binning on experimental data: We compare the proposed scheme against the low-rank approach and the SToRM: SENSE method in which the Laplacian matrix is estimated from SENSE reconstructions of the undersampled spiral data.  Temporal profiles are also shown for the whole acquisition. We note that the proposed scheme reduces blurring of the spatial images as well as the temporal profiles. Red dotted rectangles are used to show comparison of a cardiac cycle. In the low-rank method, the transition from the end of diastole phase to the end of systole is not as smooth as in the other two methods. The SENSE recovery of manifold method has more blurring as compared to the proposed method. }
	\label{fig4}
\end{figure}

\begin{figure}[htb]
	\center
	\vspace{0cm}
	\hspace{-0cm}
	\includegraphics[width=0.8\textwidth]{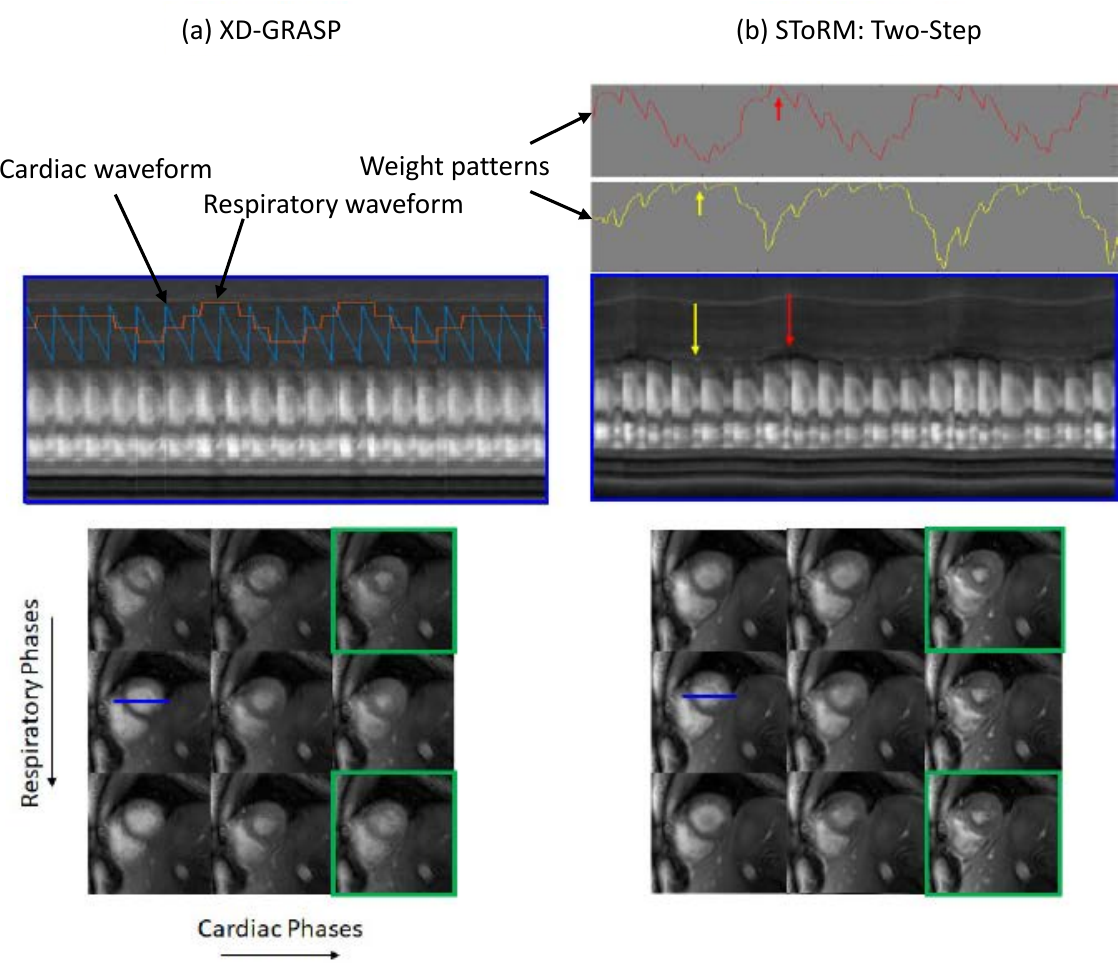}\\
	\caption{\small \textcolor{black}{Comparison of (a) XD-GRASP against (b) SToRM:Two-Step. For direct comparison of the methods, we rearrange the images obtained using SToRM into respiratory and cardiac phases (bottom-right), identified by the XD-GRASP binning approach, which can be compared to XD-GRASP reconstructions (bottom-left).   We also rearrange the XD-GRASP recovered frames to form a temporal profile in the top row. Specifically, we construct a time series by selecting the XD-GRASP frames corresponding to the identified cardiac and respiratory phases. We observe that some of the cardiac/respiratory phases are not well sampled in XD-GRASP due to variability in  the breathing cycles, resulting in blurring and aliasing artifacts. Please see the phases outlined by green boxes. By contrast, our soft-binning strategy exploits the similarity between the phases along the time series to reduce these artifacts. The weight patterns for the two frames indicated by the yellow and red arrows are shown in the top row. We note that the weights are high whenever the frames are similar to the chosen frame; the algorithm combines the information in these similar frames to obtain high-resolution reconstructions.We note that XD-GRASP is binning the data to different cardiac and respiratory bins. The averaging of motion within the bins may cause respiratory blurring, which may be the reason for difference in the hepatic vasculature. By contrast, the soft-binning offered by the proposed scheme minimizes the respiratory blurring, thus offering more sharper reconstructions.}}	
	\label{fig5}
\end{figure}

\begin{figure}[b!]
	\vspace{-0.1cm}
	\centering
	\includegraphics[ width=0.3\textwidth]{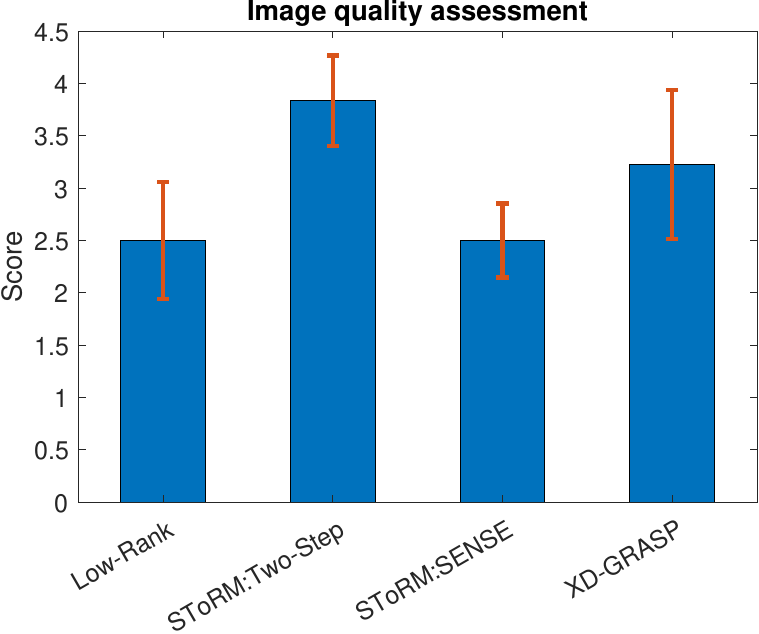}\\
	\caption {\small {Comparison of the proposed SToRM: Two-Step method against the existing methods. Image quality assessment is done in a blinded fashion by two experts on single slice data sets. Low-rank and SToRM: SENSE methods have mean scores of $2.5 \pm 0.55$ and $2.5 \pm 0.43$ respectively. Mean score of XD-GRASP = $3.3\pm 0.7$. This shows that XD-GRASP gives good results when respiratory bins are sufficiently sampled. SToRM: Two-Step mean score= $3.83 \pm 0.43$.}}
	\label{fig6}
\end{figure}

\begin{figure}
	\vspace{-0.1cm}
	\centering
	\includegraphics[width=0.45\textwidth]{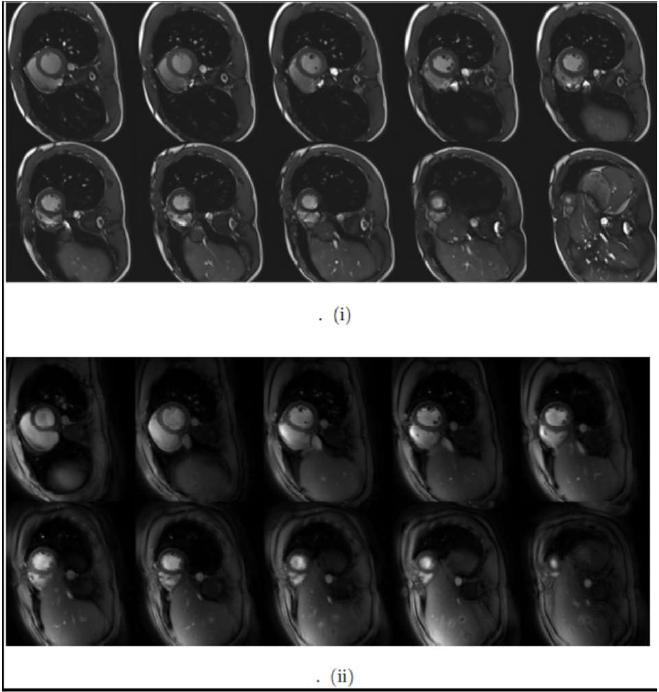}\\
	\caption {\small \textcolor{black}{Visual comparison of the whole-heart data recovered by SToRM:Two-Step against breath-held CINE data. (i) First two rows show the Cartesian SSFP based breath-held results. (ii) Second two rows show the GRE based SToRM:Two-Step reconstruction results. The increased TR and flip angle allows us to get improved GRE contrast, resulting from the inflow enhancement of the LV blood pool. The end-diastole frames are shown across all slices. Most slices of SToRM:Two-Step reconstruction show good agreement with breath-held images in terms of image quality. We stress that the direct comparison of the two methods is challenging since the two sequences differ significantly in several aspects, including image contrast and bias fields; it is difficult to draw a strong conclusion on the equivalence between the two methods, beyond the limited qualitative comparisons in Table \ref{table2}. }}
	
	\label{fig7}
\end{figure}

\section{RESULTS}

The illustrations in Fig. \ref{fig1} show the benefit of the iterative strategy in SToRM:Two-Step. The first iteration shows the SToRM: SENSE, where the Laplacian is estimated from SENSE reconstructions using the central k-space regions. Full-resolution reconstruction is obtained using this Laplacian and all of the k-space data. We observe that this results in residual aliasing artifacts. With the SENSE reconstruction and the Laplacian estimated from it as the initial guess, the kernel low-rank algorithm is run with different numbers of iterations, as shown in Fig. \ref{fig1} on the central k-space regions. The Laplacian matrix estimated from these iterations, was used to obtain the high resolution reconstruction. We note that the image quality improves significantly with iterations. In particular, the Laplacian estimated from the fifth iteration yields improved reconstructions with reduced artifacts in the liver regions and minimal myocardial blurring.  

We first compare SToRM:Two-Step, and SToRM: Self Nav with breath-held cine in Fig. \ref{fig2} to determine if the performance of the matrix completion scheme is comparable to the setting with navigators and breath-held data. The data was acquired with navigated acquisition, where one spiral acquisition was repeated after every 5 readouts. The navigators were not included in the estimation of the Laplacian matrix in the SToRM:Two-Step. We manually identified a cardiac cycle from the SToRM reconstructions, which closely matched in the end inspiration phase in which breath-held data was acquired. Three frames (end-diastole, mid-frame, end-systole) from the image series are shown. We note that the proposed scheme provides similar visual quality to the breath-held acquisitions. The experiments also show that the visual quality of the SToRM:Two-Step scheme is quite comparable to that of StoRM: Self Nav. 

We compare the proposed scheme against competing methods on a numerical phantom in Fig. \ref{fig3} and Table \ref{table1}. Fig. \ref{fig3} shows the visual comparison between the proposed SToRM:Two-Step, low-rank, SToRM: SENSE, compressed sensing (CS), kt-SLR and SToRM: Self-Nav methods. We observe that the proposed scheme significantly reduces the spatial and temporal blurring compared to the low-rank scheme CS and kt-SLR, which demonstrates the ability of the kernel low-rank algorithm in capturing non-linear redundancies. These visual observations are also confirmed by the quantitative results in Table \ref{table1}. We have used four metrics (SER, SSIM, HFEN, GPC) to evaluate the performance. We optimized the parameters for one dataset. We observe that the performance of the SToRM:Two-Step scheme is significantly better than other methods, while it is marginally higher than SToRM:Self-Nav.

We have also investigated the impact of different Laplacian estimation strategies as shown in the supplementary figure S1. Results show that the image quality is not different notably, however, we get lower computational complexity with the low-resolution approach, as mentioned in the section \ref{twostep}.  

We qualitatively compare the proposed algorithm with competing methods on six single-slice experimental datasets in Fig \ref{fig4}-\ref{fig6}. The visual comparison  against the low-rank and the SToRM:SENSE algorithm on one of the datasets is shown in Fig \ref{fig4}. We have picked three frames (end-diastole, mid-frame, end-systole) from the image series to show the spatial quality of the proposed method as compared to the low-rank method and SToRM: SENSE. The fourth column of Fig. \ref{fig4} shows the temporal profiles. We observe that the proposed scheme provides improved visual quality with reduced spatial and temporal blurring, when compared to the low-rank method and SToRM: SENSE scheme. The comparison of the proposed scheme against the XD-GRASP algorithm on the same dataset is shown in Fig. \ref{fig5}.  Since XD-GRASP and the proposed scheme use different reconstruction methodologies, we illustrate the results in two ways.  In Fig. \ref{fig5}, we display the reconstructed images in the different cardiac and respiratory phases in the bottom rows, which are identified by XD-GRASP. We create a cine movie by picking each image in the time series from the XD-GRASP reconstructions depending on the specified cardiac and respiratory phase; the time profile in the top row corresponds to a cut along the myocardium identified by the blue line in one of the images. We also display the rows of the weight matrix $\mathbf W$ corresponding to two frames, identified by the yellow and red arrows in the top right columns. As discussed earlier, the weights indicate similarity of the specific frame with other frames in the dataset. The top row in Fig. \ref{fig5}.a shows the temporal profile of XD-GRASP, while the corresponding temporal profiles of the proposed scheme are shown in Fig. \ref{fig5}.b. We note that the cardiac and respiratory phases identified by XD-GRASP are roughly in agreement with the motion patterns in the temporal profiles of the proposed scheme in Fig. \ref{fig5}.b. However, the motion patterns in the temporal profiles in XD-GRASP appear attenuated. We also observe sharp transitions in contrast between frames from different cardiac/respiratory phases. The top rows of Fig. \ref{fig5}.(b) show the weights corresponding to two frames in end inspiration and end expiration, respectively. Note that the weight patterns agree reasonably well visually with the identified cardiac respiratory phases. The weights indicate soft-binning of the phases offered by the proposed scheme. The bottom rows of Fig. \ref{fig5} show the reconstructed images arranged in the cardiac and respiratory phases, which were identified using the self-gating strategy in XD-GRASP. We note that similar binning can be performed using the eigenvectors of the Laplacian matrix as shown by Poddar et al. \cite{sunritatci}.
Fig. \ref{fig5}.a shows the recovered images using XD-GRASP, while Fig. \ref{fig5}.b corresponds to the proposed method. We note that some of the phase images are blurred in the XD-GRASP reconstructions. These phases correspond to the poorly sampled cardiac and respiratory phases. The soft-binning offered by the weighting strategy allows for more data-sharing between the phases, resulting in reduced myocardial blurring and improved fidelity of the temporal profiles. 

In Fig. \ref{fig6}, we show the image quality scores of different methods, which are rated by two experts in a blinded fashion. These results show that the image quality of SToRM:Two-Step is better than competing methods such as XD-GRASP and low-rank. We compare the scores using ANOVA with a p value of $p=0.0001$, which show that  SToRM:Two-Step and XD-GRASP are significantly different from low-rank and SToRM:SENSE, while the improvement in quality score of SToRM:Two-Step over XD-GRASP is not statistically different. However, both the raters consistently scored SToRM:Two-Step over XD-GRASP, except one expert on one of the datasets.

We compare the whole-heart breath-held SSFP data and SToRM:Two-Step on five whole heart datasets in Fig. \ref{fig7} and Table \ref{table2}. Fig. \ref{fig7}.(i) shows ten slices of SSFP based breath-held data, from base to apex of the heart, while (ii) shows the corresponding free-breathing SToRM:Two-Step reconstructions. The image quality scores as well as ejection fraction calculations, by two experts in a blinded fashion, are reported in Table \ref{table2}. The quality scores in Table \ref{table2} shows that the image quality of the proposed free-breathing strategy is slightly lower than breath-held SSFP. An ANOVA procedure with $p=0.07$ revealed that two scores are not significantly different. The ejection fraction estimated from the proposed SToRM:Two-Step scheme closely matches the breath-held SSFP method. A blocked ANOVA test showed no significant difference between breath-held and SToRM:Two-Step results ($p=0.93$), showing the accuracy of LVEF using our proposed method.

\begin{table}[h]
	\vspace{0.1cm}
	\centering
	\caption{\small Image quality assessment is done in a blinded fashion by two experts on five whole heart datasets (Both breath-held and SToRM: Two-Step reconstructed images). Second column shows the Left ventricular ejection fraction (LVEF) comparison between breath-held and SToRM: Two-Step results, calculated by two experts. Free-breathing LVEF is slightly over-estimated as compared to the breath-held LVEF.}.
	\label{table2}
	\begin{tabular}{|c|c|c|}
		\hline
		Method & Image Quality & LVEF (ml)\\ \hline
		Breath-held  & ${4.6 \pm 0.32}$    & ${57.15 \pm 2.1}$ \\ \hline
		SToRM: Two-Step&  ${3.95 \pm 0.38}$    & ${58.02 \pm 2.5}$ \\ \hline
	\end{tabular}
\end{table}
\begin{table}[h]
	\vspace{-0.1cm}
	\centering
	\caption{\small In this experiment, we have fixed the number of iterations=5 and compared all approaches using Signal to error ratio (SER) and reconstruction time. First row corresponds to the reconstructions from the single-step SToRM approach, described in Section \ref{irls}, while the two step strategy described in Section \ref{twostep} is shown in the second row. However, the computational complexity of this approach is 3-4 times higher. In addition, the difference in SER is not significant. In favor of faster experiments, we resort to two-step recovery scheme, where a low-resolution reconstruction is used to estimate the Laplacian matrix as described in Section \ref{twostep}. We have also compared Laplacian matrices estimated from original images and reconstructed images. As noted in the last row of the Table, we note that the SER of the original images is marginally higher.}
	\label{table3}
	\begin{tabular}{|c|c|c|c|}
		\hline
		Laplacian Estimation Scheme & Iterations & SER &time(minutes) \\ \hline
		Low resolution& $5$    & $25.4$ & $8$ \\ \hline
		Full resolution&  $5$    & $25.8$ & $33$ \\ \hline
		Original Images&  $0$    & $26.1$ & $3.2$ \\ \hline
	\end{tabular}
\end{table}

\begin{table}[h]
	\vspace{-0.0cm}
	\centering
	{\color{black}{
	\caption{\small {\color{black}{Shows the quantitative improvement in the reconstruction results as we increase the number of iterations. If the difference of our current update from its previous iteration is less than threshold ($1e^{-6}$) or it reaches maximum number of iterations, we stop our iterative scheme.}}}.
	\label{table4}
	\begin{tabular}{|c|c|c|c|c|c|}
		\hline
		Iteration number & Iter 1 & Iter 2 & Iter 3 & Iter 4 & Iter 5 \\ \hline
		SER & $17.40$    & $19.40$ & $22.80$ & $24.75$ & $25.40$ \\ \hline
	\end{tabular}
}}
\end{table}

\section{DISCUSSION}
We introduced an iterative spiral-SToRM framework for the recovery of free-breathing and ungated cardiac images from 2-D spiral acquisition. The framework assumes the images to be on a smooth surface in high dimensions and relies on a kernel low-rank prior to recover the dataset. {\textcolor{black}{The main difference of our scheme from our prior work \cite{storm} is the use of an iterative kernel low-rank matrix completion algorithm and we are using $l_2$ smoothness regularization as it is giving us better results as compared to the $l_1$ smoothness regularization. The proposed approach eliminates the need for explicit k-space navigators and relies on variable-density spiral acquisitions, where the central k-space regions are acquired with higher sampling density. By eliminating the need for navigators, the proposed scheme improves sampling efficiency and hence image quality. The use of spiral trajectory improved sampling efficiency and contrast. Specifically, the increased TR and increased flip angle offer improved contrast compared to the low TR radial acquisitions.} To improve computational efficiency, we rely on a two-step strategy. In the first step, we estimate low-resolution reconstructions as well as the Laplacian matrix from the central k-space region using a kernel low-rank optimization scheme. Once the Laplacian matrix is estimated, we solve for the high-resolution image from the entire k-space data using the manifold Laplacian. 
The benefit of using low resolution approach is shown in the Table \ref{table3} and in the supplementary figure. We have compared low and high resolution approaches using SER and processing time. Low resolution reconstruction reduces the computational complexity significantly.
We also approximate the Laplacian using a few basis functions, which reduces the computational complexity and memory demand of the algorithm by an order of magnitude. We observe that the SToRM: Two-Step approach recovers 2D cine images with reduced spatial and temporal blurring in a short free-breathing self-gated acquisition, compared to low-rank and explicit binning strategies.
We have also compared Laplacian matrices estimated from original images and reconstructed images. The SER obtained from original images $=26.1$, whereas SER from the reconstructed images is $25.8$

The gradient echo (GRE) acquisition schemes have few advantages for simplifying 3T cine imaging, even though SSFP sequences are typically used for cine imaging. The longer repetition time (TR) in the spiral trajectory provides inflow-enhancement of the LV blood pool; the resulting contrast is similar to the Cartesian SSFP imaging as compared to the shorter TR Cartesian GRE imaging. Furthermore, the spoiled GRE-based approach used for the acquisition is robust to banding artifacts, which SSFP methods are vulnerable to, without any  frequency scout requirement. In addition, GRE schemes are less sensitive to eddy current artifacts caused by the large angular increment of the golden angle ordering \cite{wundrak2016golden}. With 16 seconds per slice, the whole heart is imaged in 3 minutes.  The difference in blood pool-myocardium contrast between diastole and systole seen in  SToRM:Two-Step are due to inflow effects associated with GRE acquisitions. 


Our experiments in Fig. \ref{fig5} show that the proposed scheme provides less-blurred reconstructions compared to XD-GRASP. As discussed previously, XD-GRASP relies on binning each image to appropriate cardiac/respiratory phases. We note that the time duration of heart in all of the cardiac/respiratory phases is not equal, with some phases (e.g. mid-systolic and inspiration phases ) having significantly fewer spokes than others. The recovery of these images from a few k-space points is significantly more challenging in XD-GRASP, which results in the residual blurring. By contrast, the SToRM strategy relies on a soft-gating strategy with no explicit binning. Our experiments in Fig. \ref{fig5} show that this approach is more robust to residual blurring. 

Our quantitative and qualitative experiments show that the SToRM:Two-Step scheme provides reconstructions that are marginally better in quality than SToRM:Self-Nav in the simulation settings. While both methods rely on the kernel low-rank algorithm, SToRM:Two-Step scheme does not require specialized k-space navigators to estimate the manifold Laplacian. The major benefit of this scheme is its application to golden angle radial or spiral sequences already in place on several scanners.  In this work, we have not compared the two approaches on experimental MR data. Specifically, the experimental spiral data was acquired without any navigators, which makes it difficult to realize SToRM:Self-Nav. Thus, based on the current experiments, we cannot conclude that SToRM: Two-Step is superior to SToRM: Self-Nav; we plan to pursue these comparisons in our future work.

The comparisons with breath-held SSFP data in Fig. \ref{fig7} and Table \ref{table2} show that the image quality and ejection fraction measures of the proposed scheme are comparable to breath-held acquisitions. However, the two sequences differ significantly in several aspects, including image contrast and bias-fields. We notice minor blurring of the myocardium in the slices close to the apex. This problem may be due to the inaccurate estimation of cardiac motion patterns from the navigators. Specifically, when the heart occupies a tiny fraction of the field of view as with apical slices, the in-plane navigators may not be sensitive to cardiac motion. In the future, we propose to extend SToRM with 3D navigators to minimize the above problem.

\vspace{-0.1cm}
\section{Conclusion}
In this paper, we have proposed an iterative SToRM algorithm (SToRM:Two-Step) for the recovery of free-breathing and ungated cardiac MR images using spiral acquisitions with no k-space navigators. Our experiments show that the proposed scheme offers better performance compare to our previous STORM:Self-Nav method, which requires k-space navigators.}} Qualitative comparisons by experts also show that the proposed scheme provides less spatial and temporal blurring compared to low-rank methods, which do not require explicit binning to cardiac/respiratory phases, and XD-GRASP, which bins the data. Our preliminary study also shows that the proposed framework provides ejection fraction measures, which are statistically equivalent to the breath-held measurements.
The MATLAB code for the SToRM-Iterative scheme can be downloaded from\\ https://github.com/ahaseebahmed/SpiralSToRM-Iterative.

\vspace{-0.1cm}

\end{document}